\documentclass[aps,prb,twocolumn,superscriptaddress,showpacs]{revtex4}
\makeatletter

\newcommand{\Rmnum}[1]{\expandafter\@slowromancap\romannumeral #1@}
\makeatother

\usepackage{eurosym}
\usepackage{amsfonts}
\usepackage{amssymb}
\usepackage{amsmath}
\usepackage{graphicx}
\usepackage{color}
\begin{document}

\title{Perfect charge compensation in WTe$_{2}$ for the extraordinary magnetoresistance: From bulk to monolayer}

\author{H. Y. Lv}

\affiliation{Key Laboratory of Materials Physics, Institute of Solid State Physics, Chinese Academy of Sciences, Hefei 230031, People's Republic of China}

\author{W. J. Lu}
\email[Corresponding author: ]{wjlu@issp.ac.cn}
\affiliation{Key Laboratory of Materials Physics, Institute of Solid State Physics, Chinese Academy of Sciences, Hefei 230031, People's Republic of China}

\author{D. F. Shao}

\affiliation{Key Laboratory of Materials Physics, Institute of Solid State Physics, Chinese Academy of Sciences, Hefei 230031, People's Republic of China}

\author{Y. Liu}

\affiliation{Key Laboratory of Materials Physics, Institute of Solid State Physics, Chinese Academy of Sciences, Hefei 230031, People's Republic of China}

\author{S. G. Tan}

\affiliation{Key Laboratory of Materials Physics, Institute of Solid State Physics, Chinese Academy of Sciences, Hefei 230031, People's Republic of China}

\author{Y. P. Sun}
\email[Corresponding author: ]{ypsun@issp.ac.cn}
\affiliation{Key Laboratory of Materials Physics, Institute of Solid State Physics, Chinese Academy of Sciences, Hefei 230031, People's Republic of China}
\affiliation{High Magnetic Field Laboratory, Chinese Academy of Sciences, Hefei 230031, People's Republic of China}
\affiliation{Collaborative Innovation Center of Advanced Microstructures, Nanjing University, Nanjing, 210093, People's Republic of China}

\makeatletter


\begin{abstract}
The electronic structure of WTe$_{2}$ bulk and layers are investigated by using the first principles calculations. The perfect electron-hole ($n$-$p$) charge compensation and high carrier mobilities are found in WTe$_{2}$ bulk, which may result in the large and non-saturating magnetoresistance (MR) observed very recently in the experiment [Ali $et$ $al$., Nature \textbf{514}, 205 (2014)]. The monolayer and bilayer of WTe$_{2}$ preserve the semimetallic property, with the equal hole and electron carrier concentrations. Moreover, the very high carrier mobilities are also found in WTe$_{2}$ monolayer, indicating that the WTe$_{2}$ monolayer would have the same extraordinary MR effect as the bulk, which could have promising applications in nanostructured magnetic devices.
\end{abstract}

\maketitle

In the past few decades, transition metal dichalcogenides (TMDs) with the formula MX$_{2}$ (M=transition metal of group 4-10; X=S, Se, or Te) have received a lot of attention due to their diverse properties. Ranging from insulating to metallic, some of them exhibit behaviors of superconductivity\cite{B.Sipos-NatureMater} and charge density wave.\cite{J.A.Wilson-1975} Most of the MX$_{2}$ systems crystallize in a layered structure, with the building blocks X-M-X sandwiches stacked along the $c$-axis. Within the sandwich layer, the atoms are covalently bonded, while the interactions among the layers are much weaker, mainly of the van der Waals type. The recent advances in the experimental techniques have made it possible to exfoliate two-dimensional (2D) ultrathin layers from the MX$_{2}$ bulk.\cite{synthesis1,synthesis2} The 2D MX$_{2}$ materials can largely preserve the versatile properties, but some distinctive characteristics can be introduced due to the quantum confinement effect. For example, the transition of indirect-direct band gap takes place in MoS$_{2}$ when the bulk structure is exfoliated into a monolayer,\cite{quantum-confine} which opens up new potential applications in the fields of phototransistors,\cite{phototransistor} photocatalyst,\cite{photocatalyst} electroluminescence,\cite{electroluminescence} etc.

Very recently, extremely large magnetoresistance (MR) without saturation even at very high fields was observed in WTe$_{2}$.\cite{Cava-Nature} MR evaluates the change in electrical resistance by the application of a magnetic field. Large MR effect can have promising applications such as magnetic field sensors\cite{Sensors} and magnetic information storage.\cite{Prinz-Science} The main origin of the extraordinary MR effect in WTe$_{2}$ is ascribed to the perfect $n$-$p$ charge compensation in this material, based on the investigations by the angle-resolved photoelectron spectroscopy (ARPES).\cite{Cava-PRL} When applied in nanoelectronics, it is desirable to obtain nanostructured systems which have the performance as good as or even better than their bulk counterparts. Almost at the same time, huge negative MR effect was reported in the ultrathin layers of TiTe$_{2-x}$I$_x$ due to the frustrated magnetic structures induced by the anionic doping.\cite{Xieyi-PRL} The excellent MR effect of WTe$_{2}$ bulk inspires us to explore how the WTe$_{2}$ layer will perform, which is very critical in the applications of nanoelectronics.

In this work, the electronic properties of WTe$_{2}$ layers as well as the bulk structure are investigated based on the first-principles calculations. Our results show that both the monolayer and bilayer of WTe$_{2}$ maintain the same semimetal properties as the bulk, with equal hole and electron carrier concentrations, suggesting that the non-saturating MR effect may also exist in the WTe$_{2}$ layers. Moreover, the high carrier mobilities found in WTe$_{2}$ monolayer indicate that the MR effect in the monolayer would be comparable to that in the bulk system.

\begin{figure}
\includegraphics[width=1.0\columnwidth]{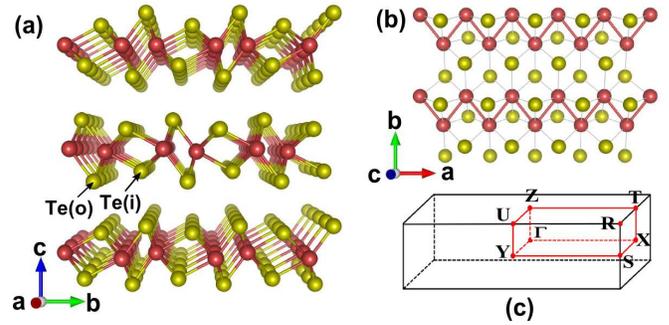}\caption{\label{fig1-crystal structure} Crystal structure of WTe$_{2}$ bulk viewed along (a) the $a$ axis (parallel to the W-W zigzag chains) and (b) the $c$ axis (perpendicular to the stacked layers); (c) the corresponding first Brillouin zone. The red and yellow balls represent W and Te atoms, respectively. Te(i) and Te(o) stand for the atoms shrunk inside and moved outside the sandwich layer, respectively.}
\end{figure}

Our calculations were performed $via$ a projector augmented wave (PAW) pseudopotential approach within the density functional theory (DFT) as implemented in the ABINIT code.\cite{X.Gonze-2009,X.Gonze-2002,X.Gonze-2005} The generalized gradient approximation (GGA) with the Perdew-Burke-Ernzerhof (PBE) functional\cite{PBE-1996} were used for exchange-correlation energy. For bulk and bilayer, the van der Waals interactions were treated by the vdW-DF1 functional.\cite{vdW-Dion} Spin-orbit coupling was included in the calculations of the electronic properties. The plane-wave cutoff energy was set to be 600 eV in all the calculations. In the self-consistent calculations, the Brillouin zones were sampled with an $8\times4\times1$ Monkhorst-Pack $k$ mesh. All the structures were fully relaxed until the force acting on each atom became less than $0.5 \times 10^{-3}\,{\mbox{eV}}/{\mbox{\AA}}$.

The crystal structure of the WTe$_{2}$ bulk viewed along different directions are demonstrated in Figs. 1(a) and (b), respectively, with the corresponding first Brillouin zone shown in Fig. 1(c). The space group of WTe$_{2}$ bulk is $P$nm2$_{1}$. Most of the sulfides and selenides are rhombohedral or hexagonal, and the metal atoms are trigonal prismatic or octahedral coordinated by six chalcogen atoms. However, in WTe$_{2}$, the octahedron of tellurium atoms is slightly distorted and the metal atoms are displaced from their ideal octahedral sites, forming zigzag metal-metal chains along the $a$ axis,\cite{A.Mar-Jacs} which is demonstrated in Fig. 1(b). The structural difference has endowed this kind of material with properties distinct from the other MX$_{2}$ system. After fully relaxation, the obtained crystal constants are $a$=3.54 {\mbox{\AA}}, $b$=6.34 {\mbox{\AA}} and $c$=14.44 {\mbox{\AA}}, which agree well with the experimental results.\cite{A.Mar-Jacs} The nearest distance between W-W atoms along the zigzag chain is 2.87 {\mbox{\AA}}, only 0.14 {\mbox{\AA}} larger than that in the pure metal crystal of tungsten. Because of the distorted octahedral structure, the Te atom layers become buckled (see Fig. 1(a)), with Te(i) atoms shrunk a little inside the sandwich layer and Te(o) atoms moved slightly outside.

\begin{figure}
\includegraphics[width=1.0\columnwidth]{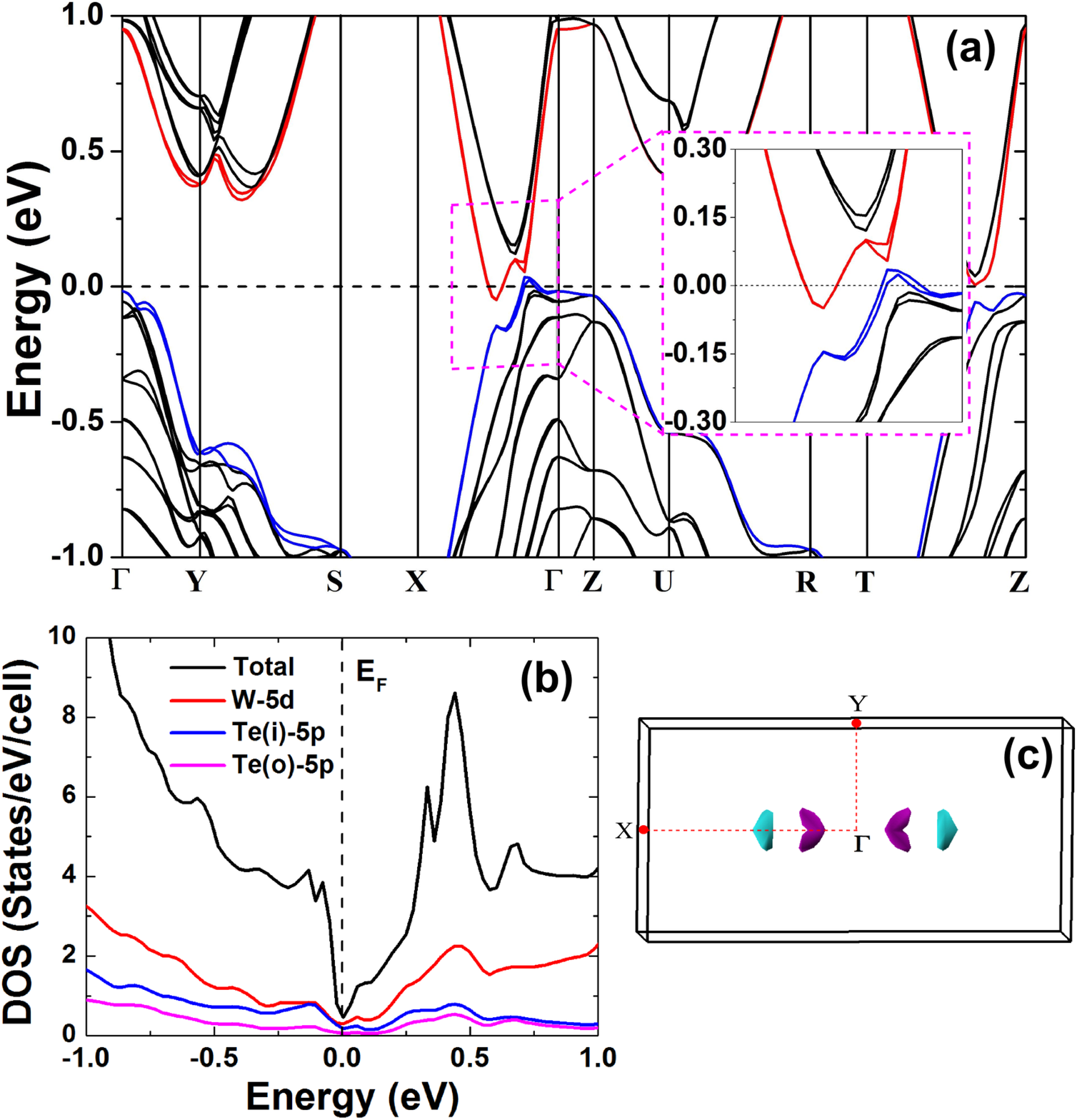}\caption{\label{fig2-band structure of WTe$_{2}$ bulk}(a) Band structure, (b) total and partial density of states (DOS), and (c) Fermi surface (FS) of WTe$_{2}$ bulk.}
\end{figure}

The electronic band structure and the density of states (DOS) of WTe$_{2}$ bulk are shown in Figs. 2(a) and (b), respectively. A strong anisotropic band dispersion is observed in the band structure. The bands along the $\Gamma$-$Z$ direction (perpendicular to the sandwich layers in real space) are much flatter than those along the other directions, reflecting the quasi-two-dimensional layered structure of WTe$_{2}$ bulk. The very small DOS at the Fermi energy ($E_F$) signals the semimetallic nature of WTe$_{2}$ bulk. The corresponding W-5$d$ and Te-5$p$ partial DOS included in Fig. 2(b) show that the DOS at the $E_F$ ($N(E_F)$) is dominated by the W-5$d$ state, followed by the Te(i)-5$p$ state. The Te(o)-5$p$ state contributes little to the $N(E_F)$. The Fermi surface of WTe$_{2}$ bulk is demonstrated in Fig. 2(c), which exhibits highly anisotropic property. Note that in the band structure, there exists a small overlap between the top of the valence band (hole pocket) and bottom of the conduction band (electron pocket) along the $X$-$\Gamma$ direction in the vicinity of the Fermi level. The enlarged overlap part of the band structure is shown in the inset of Fig. 2(a). Hole and electron pockets with appropriately the same size are found, which agrees well with the ARPES results.\cite{Cava-PRL} In experiment, it was analyzed that these two pockets may lead to the perfect carrier compensation and therefore the large MR effect in WTe$_{2}$. However, the exact values of the hole ($p$-type) and electron ($n$-type) carrier concentrations are still lacking. To obtain these values, the band-decomposed DOSs are calculated. In particular, to calculate the $p$-type carrier concentration, we calculate the DOS of the bands in blue color in Fig. 2(a) and then integrate the DOS from $E_F$ to the valence band maximum (VBM); to calculate the $n$-type carrier concentration, the DOS of the bands in red color is integrated from the $E_F$ to the conduction band minimum (CBM). The carrier concentrations, $p=7.7\times10^{19} \mbox{cm}^{-3}$ and $n=7.5\times10^{19} \mbox{cm}^{-3}$, are obtained, coinciding perfectly with each other.

\begin{table*}

\caption{\label{Table1-carrier mobility} Effective mass $m^*$, elastic modulus $C$, deformation potential constant $E_1$, and room-temperature carrier mobility $\mu$ of WTe$_{2}$ bulk and monolayer.}

\begin{tabular}{ccccccccccccc}
\hline
\hline
~&~& direction &~& carrier type &~& $m^*$ &~& $C$ &~& $E_1$ &~& $\mu(\times10^3)$\tabularnewline
~&~& &~& &~& ($m_0$) &~& eV/${\mbox{\AA}}^3$(eV/${\mbox{\AA}}^2)$ &~& eV &~& $\mbox{cm}^2\mbox{V}^{-1}\mbox{s}^{-1}$\tabularnewline
\hline
bulk &~& $a$-axis &~& hole &~& 0.19 &~& 0.94 &~& $-$7.24 &~& 11.23\tabularnewline
~&~& &~& electron &~& 0.15 &~& 0.94 &~& $-$4.90 &~& 44.16\tabularnewline
monolayer &~& $a$-axis &~& hole &~& 0.89 &~& 7.72 &~& $-$10.41 &~& 0.02\tabularnewline
~&~& &~& electron &~& 0.30 &~& 7.72 &~& $-$4.27 &~& 1.09\tabularnewline
~&~& $b$-axis &~& hole &~& 0.54 &~& 9.12 &~& 0.58 &~& 21.25\tabularnewline
~&~& &~& electron &~& 0.28 &~& 9.12 &~& $-$1.42 &~& 13.13\tabularnewline
\hline
\hline
\end{tabular}

\end{table*}

In the semiclassical two-band model,
\begin{equation}\label{1}
  MR=\frac{\sigma\sigma^\prime(\sigma/n+\sigma^\prime/p)^2(B/e)^2}{(\sigma+\sigma^\prime)^2+\sigma^2{\sigma^\prime}^2(1/n-1/p)^2(B/e)^2},
\end{equation}
where $\sigma$ and $\sigma^\prime$ are the electrical conductivities of electrons and holes without external magnetic field, respectively. $n$ and $p$ are the electron and hole concentrations, respectively. When $n=p$, the MR increases as $B^2$ without saturation. Using this two-band model, we can qualitatively interpret the experimentally observed behavior of MR as a function of external magnetic field in WTe$_{2}$.

Next, we focus on the electronic properties of WTe$_{2}$ ultrathin layers. The calculated electronic structures of WTe$_{2}$ monolayer and bilayer are plotted in Figs. 3(a) and (b), respectively. What is different from the case of WTe$_{2}$ bulk is that the VBM of WTe$_{2}$ layers are located at the $\Gamma$ point. From monolayer to bilayer, more valence (in blue color) and conduction (in red color) bands cross the Fermi energy, thus the overlap of valence and conduction bands becomes larger in bilayer, which can also be seen from the Fermi surface in Figs. 3(e) and (f). Combined with the calculated DOS (see Figs. 3(c) and (d)), we can see that the WTe$_{2}$ monolayer and bilayer remain semimetals. The semimetallic property is very different from that reported in the WTe$_{2}$ monolayer with the artificial $2H$ structure,\cite{Amin-RSCadvance} which is a direct-band-gap semiconductor. The $N(E_F)$ for both the monolayer and bilayer are dominated by the W-5$d$ state, followed by the Te(i)-5$p$ state, which is the same as the case of WTe$_{2}$ bulk. Interestingly, for the semimetal monolayer and bilayer, equal $n$- and $p$-type carrier concentrations are also obtained. The calculated carrier concentrations for monolayer and bilayer are, respectively, $n=p=1.6\times10^{13} \mbox{cm}^{-2}$ and $n=p=1.4\times10^{13} \mbox{cm}^{-2}$. The perfect charge compensation indicates that the non-saturating MR as a function of the external magnetic field may also exist in the WTe$_{2}$ monolayer and bilayer.

\begin{figure}
\includegraphics[width=1.0\columnwidth]{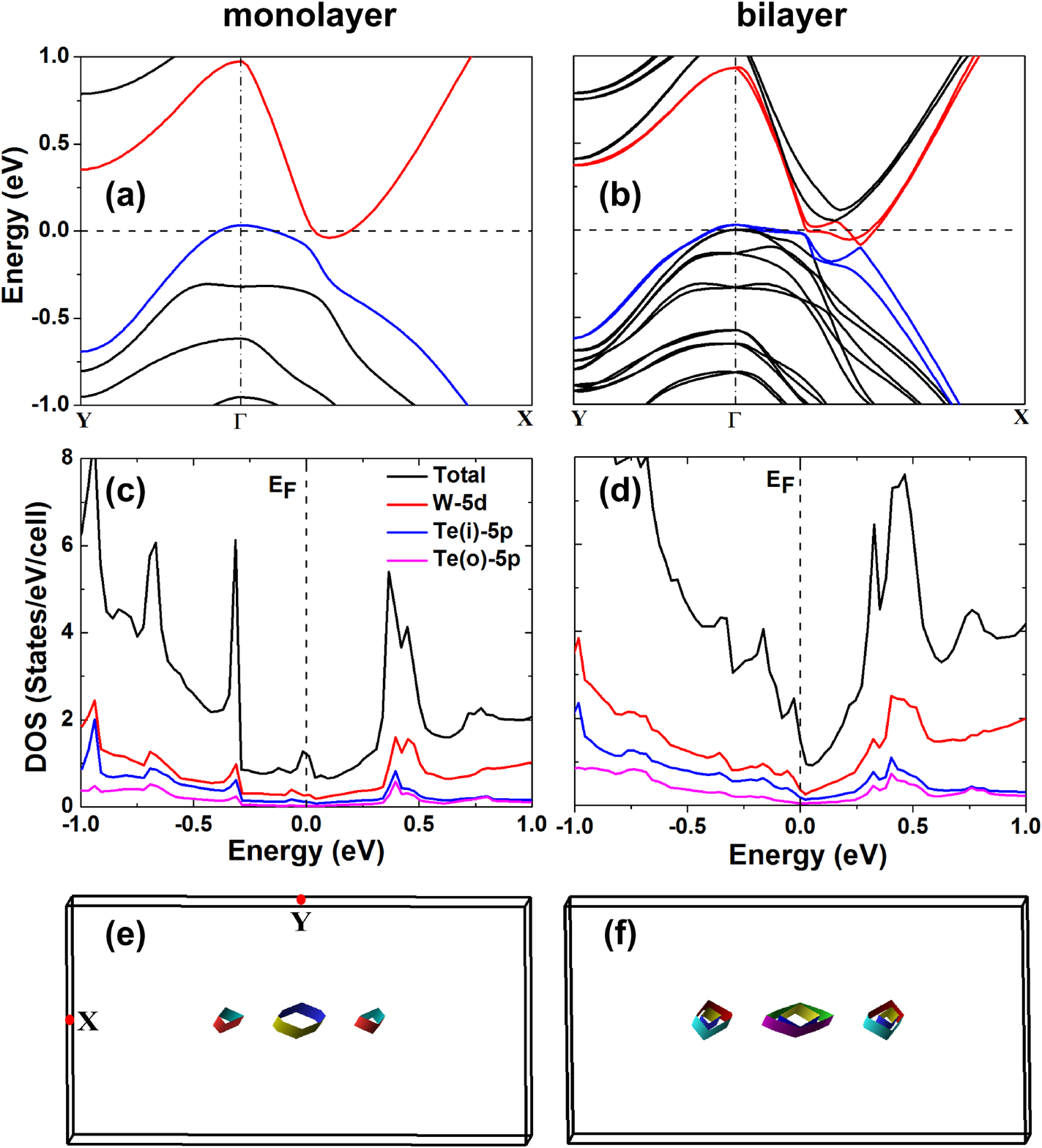}\caption{\label{fig3-band structure of WTe$_{2}$ layer} Band structure of WTe$_{2}$ (a) monolayer and (b) bilayer; total and partial density of states (DOS) of WTe$_{2}$ (c) monolayer and (d) bilayer; Fermi surface (FS) of WTe$_{2}$ (e) monolayer and (f) bilayer.}
\end{figure}

From equation (1), we can see that when $n=p$, $MR$ obeys $MR=\sigma\sigma^\prime(B/e)^2/n^2=\mu_e\mu_h$B$^2$, where $\mu_e$ and $\mu_h$ are the carrier mobilities of electrons and holes, respectively. Therefore, in order to obtain a large MR at a specific magnetic field, high carrier mobility will be desirable. The carrier mobility can be calculated using the deformation potential (DP) model based on the effective mass approximation.\cite{J.Bardeen-1950,P.J.Price-1981,J.Xi-2012} The mobilities for the bulk ($\mu_\beta^{3D}$) and two-dimensional system ($\mu_\beta^{2D}$) along a certain direction $\beta$ are respectively expressed as
\begin{equation}\label{1}
    \mu_\beta^{3D}=\frac{2\sqrt{2\pi}e\hbar^4C_\beta^{3D}}{3(k_BT)^{3/2}(m^*)^{5/2}E_1^2}
\end{equation}
and \begin{equation}\label{1}
    \mu_\beta^{2D}=\frac{2e\hbar^3C_\beta^{2D}}{3k_BT|m^*|^2E_1^2}.
\end{equation}
Here $C$ is the elastic modulus and can be defined as $C_\beta^{3D}=[\partial^2 E/\partial\delta^2]/V_0$ and $C_\beta^{2D}=[\partial^2 E/\partial\delta^2]/S_0$ for 3$D$ and 2$D$ systems, respectively, where $E$, $\delta$, $V_0$, and $S_0$ are, respectively, the total energy, the applied strain along $\beta$ direction, the volume, and the area of the investigated system. $T$ is the temperature and $m^*$ is the effective mass. The DP constant $E_1$ is obtained by $E_1=dE_{edge}/d\delta$, where $\delta$ is the applied strain by a step of 0.5\% and $E_{edge}$ is the energy of the band edges (VBM for the holes and CBM for the electrons). Here we only compare the values of WTe$_{2}$ monolayer with those of the bulk. The calculated $m^*$, $C$, $E_1$, and room-temperature $\mu$ are summarized in Table I. For WTe$_{2}$ bulk, because the large MR was measured along the $a$ axis, we only calculate the mobilities along this direction. The calculated effective masses of the hole and electron are 0.19 $m_0$ and 0.15 $m_0$ ($m_0$ is the mass of an electron), respectively. The very small effective masses result in the extremely high carrier mobilities in WTe$_{2}$ bulk, $1.12\times10^4$ and $4.42\times10^4$ $\mbox{cm}^2\mbox{V}^{-1}\mbox{s}^{-1}$ for $p$- and $n$-type carriers, respectively. The very large $\mu_h$ and $\mu_e$ as well as the perfect $n$-$p$ charge compensation as discussed above may be responsible for the measured large and non-saturating MR in WTe$_{2}$ bulk. For WTe$_{2}$ monolayer, the effective mass $m^*$ of hole along the $a$ axis is 0.89 $m_0$, larger than that along the $b$ axis (0.54 $m_0$), due to the relatively flatter band along the $\Gamma$-$X$ direction. Besides, DP constants $E_1$ along the $a$ axis are much larger than those along the $b$ axis. Consequently, the mobilities along the $b$ axis are much larger than those along the $a$ axis. The electronic transport of  WTe$_{2}$ monolayer exhibits strong anisotropic property. On the other hand, for the direction of $b$ axis, the mobilities of both the hole and electron as well as their product ($\mu_h\mu_e$) are comparable to the bulk results. Thus, our results indicate that not only the WTe$_{2}$ bulk but also the monolayer may exhibit the extraordinary MR effect.

In summary, we have investigated the electronic properties of WTe$_{2}$ bulk and layers. The perfect charge compensation as well as the large carrier mobilities are found in WTe$_{2}$ bulk, which is ascribed to be the source of the large and non-saturating MR observed experimentally. Moreover, WTe$_{2}$ monolayer and bilayer preserve the semimetallic property and both of them are found to have equal hole and electron carrier concentrations. Our results indicate that the same extraordinary MR effect as in WTe$_{2}$ bulk may also exist in WTe$_{2}$ monolayer, which will have promising applications in nanostructured magnetic devices. Further experimental investigations are deserved to confirm our computational results.

\vspace{3ex}
This work was supported by the National Key Basic Research under Contract No. 2011CBA00111, the National Natural Science Foundation of China under Contract Nos. 11274311 and 11404340, the Joint Funds of the National Natural Science Foundation of China and the Chinese Academy of Sciences' Large-scale Scientific Facility (Grant No. U1232139), the Anhui Provincial Natural Science Foundation under Contract No. 1408085MA11, and the China Postdoctoral Science
Foundation (Grant No. 2014M550352). The calculation was partially performed at the Center for Computational Science, CASHIPS.

\end{document}